\documentclass[aps,pre,reprint,superscriptaddress,amsmath,showpacs]{revtex4-1}
\usepackage[hyperindex,breaklinks,colorlinks=true]{hyperref}
\usepackage{graphicx,sidecap,enumerate,latexsym,amsfonts,amssymb,amsmath,mathrsfs,amsfonts,romanbar,mathtools,sidecap}
\usepackage{pifont}
\usepackage{xcolor}
\usepackage{verbatim}
\usepackage{chngcntr}

\begin{document}
\title{Impact of environmental changes on the dynamics of temporal networks}
\author{Hyewon Kim}
\affiliation{Asia Pacific Center for Theoretical Physics, Pohang 37673, Republic of Korea}
\author{Hang-Hyun Jo}
\affiliation{Asia Pacific Center for Theoretical Physics, Pohang 37673, Republic of Korea}
\affiliation{Department of Physics, Pohang University of Science and Technology, Pohang 37673, Republic of Korea}
\affiliation{Department of Computer Science, Aalto University, Espoo FI-00076, Finland}
\author{Hawoong Jeong}
\email[]{hjeong@kaist.edu}
\affiliation{Department of Physics, Korea Advanced Institute of Science and Technology, Daejeon 34141, Republic of Korea}
\affiliation{Center for Complex Systems, Korea Advanced Institute of Science and Technology, Daejeon 34141, Republic of Korea}

\date{\today}

\begin{abstract}
Dynamics of complex social systems has often been described in the framework of temporal networks, where links are considered to exist only at the moment of interaction between nodes. Such interaction patterns are not only driven by internal interaction mechanisms, but also affected by environmental changes. To investigate the impact of the environmental changes on the dynamics of temporal networks, we analyze several face-to-face interaction datasets using the multiscale entropy (MSE) method to find that the observed temporal correlations can be categorized according to the environmental similarity of datasets such as classes and break times in schools. By devising and studying a temporal network model considering a periodically changing environment as well as a preferential activation mechanism, we numerically show that our model could successfully reproduce various empirical results by the MSE method in terms of multiscale temporal correlations. Our results demonstrate that the environmental changes can play an important role in shaping the dynamics of temporal networks when the interactions between nodes are influenced by the environment of the systems.
\end{abstract}

\maketitle

\section{Introduction}
Dynamical behaviors of various complex systems can be described by temporal patterns of interactions among constituents of the systems, which have recently been studied in the framework of temporal networks~\cite{Holme2012, Holme2019}. This is partly due to the increasing availability of datasets with high temporal resolution~\cite{Cattuto2010, Stehle2011, Blondel2015, Karsai2018, Bhattacharya2019}. A number of temporal interaction patterns in natural and social phenomena are found to be non-Poissonian or bursty~\cite{Karsai2018} and they have been known to strongly influence the dynamical processes taking place in the system, such as spreading and diffusion~\cite{Iribarren2009, Salathe2010, Hill2010, Karsai2011, Rocha2011, Starnini2012, Jo2014, Scholtes2014, Masuda2017}. In addition, the dynamical properties of temporal networks were studied in terms of the effects of temporal resolution and time ordering of interactions~\cite{Pfitzner2013, Ribeiro2013, Krings2012}. To understand the underlying mechanisms behind empirical findings for temporal networks, several modeling approaches have been taken: These models could successfully generate characteristics of real-world temporal networks such as heavy-tailed degree distributions, community structure, and/or bursty behaviors~\cite{Jo2011, Perra2012, Karsai2014, Medus2014, Kim2015, Moinet2015, Ubaldi2016, Li2019}, enabling us to better understand the interaction mechanisms in temporal networks.

In general, the dynamics of complex social systems is driven by both internal and external factors. The internal factors may correspond to the individual attributes or the nature of relationships between individuals. The internal factors may not be the only driving force for the bursty interaction patterns between individuals, which can also be affected by various external factors. The obvious external factors in human social behaviors are circadian, weekly, and even longer cycles~\cite{Malmgren2008, Zhou2012, Jo2012, Yasseri2012, Sun2013, Stopczynski2014, Aledavood2015, Gandica2016, Monsivais2017, Pan2017, Peixoto2018, Huang2018, Lynn2019}. Despite the importance of such external factors in understanding the temporal correlations observed in temporal networks, we find only few studies on the effects of external factors on bursty temporal interaction patterns. These effects have been studied, e.g., by modeling circadian and weekly patterns with a periodic event rate or activity level~\cite{Malmgren2008, Zhou2012} or by de-seasoning the cyclic behaviors from the bursty time series~\cite{Jo2012}. Our understanding of such effects is far from complete, which clearly calls for more rigorous and systematic studies. 

In this paper, we investigate the impact of the time-varying external factors or environmental changes on temporal correlations in temporal networks. We first analyze the several temporal network datasets, some of which are known to be affected by the time-varying external factors, by means of the multiscale entropy (MSE) method~\cite{Costa2002, Costa2005} for detecting temporal correlations in multiple timescales. This is because the time-varying external factors are expected to introduce non-trivial long-range temporal correlations in the dynamics of temporal networks. By the MSE method, we find that the datasets analyzed can be categorized according to the environmental similarity. Then we devise a temporal network model that considers both internal and external factors. Here the external factor is assumed to be periodic in time, while the internal one is constant of time. Incorporating the preferential interaction mechanism into the model, we successfully generate various patterns of temporal correlations in the temporal networks. Our modeling approach helps us better understand how the environmental changes may affect the non-trivial temporal interaction patterns observed in the empirical temporal networks.

Our paper is organized as follows. In Sec.~\ref{section2}, after briefly reviewing the MSE method, we show the results by the MSE method for empirical temporal networks. In Sec.~\ref{section3}, we devise a temporal network model that considers both internal and external factors to discuss the effects of the environmental changes on the dynamics of temporal networks. Finally, we conclude our work in Sec.~\ref{section4}.

\section{Temporal correlations in data}
\label{section2}
To characterize temporal correlations in temporal networks, we apply the multiscale entropy (MSE) method~\cite{Costa2002, Costa2005} to the network-level time series of empirical face-to-face interaction datasets. As for the network-level time series, we first consider a time series for the number of interactions between individuals or activated links, as it is the simplest quantity measuring the overall interaction patterns of the temporal network. We also study the time series of the number of newly activated links or links that are activated for the first time. This quantity may capture the evolutionary dynamics of the network topology because the first activation of a link can be interpreted as the creation of the link.

\begin{figure}[!t]
\center
\includegraphics[width=\columnwidth]{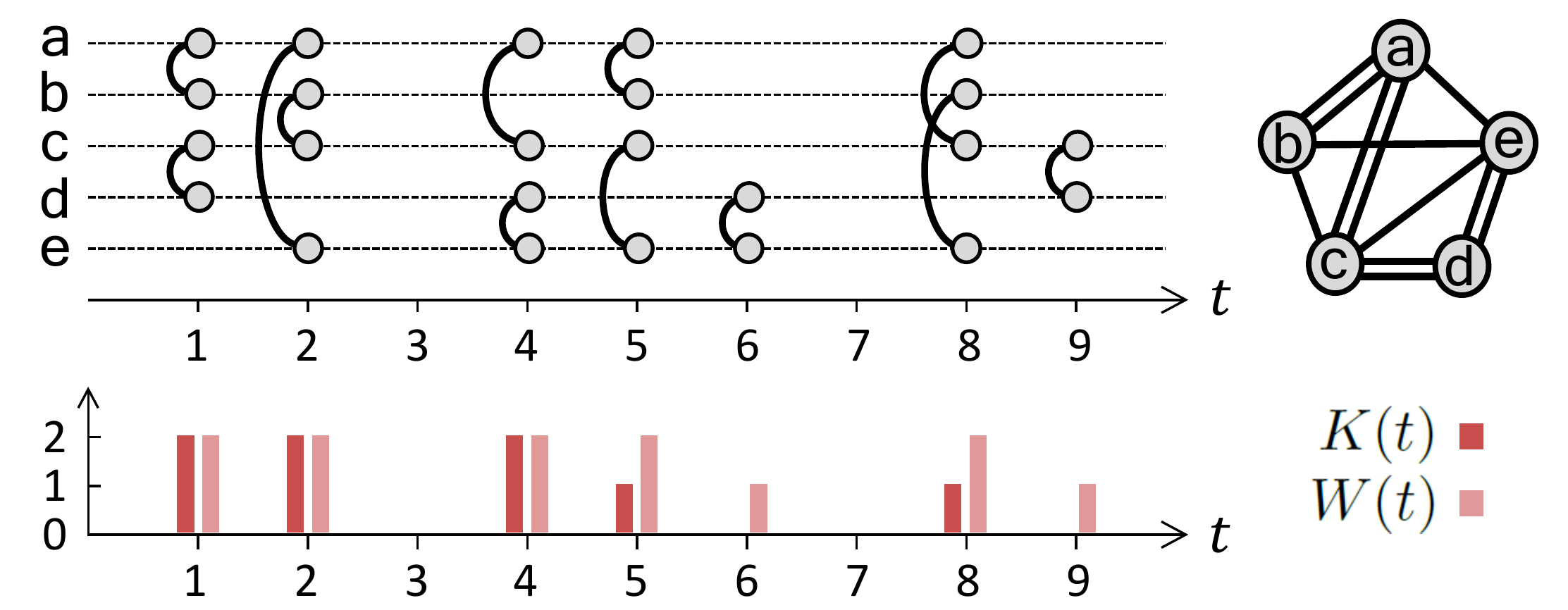}
\caption{An example of a temporal network in discrete time: the time-resolved and time-aggregated representations of the network (top) and the network-level time series of $\{K(t)\}$ and $\{W(t)\}$ (bottom).}
\label{fig1}
\end{figure}

We introduce notations for the temporal network with $N$ nodes, $K$ links, and the total number of activations over all links $W$ during the observation period of $T$. Each link $i$ ($i=1,\dots,K$) at a time step $t$ ($t=1,\dots, T$) can be either in an active state (interaction) or in an inactive state (no interaction), which are denoted by $a_i(t)=1$ and $0$, respectively. The number of activated links at the time step $t$, denoted by $W(t)$, is given as $W(t)=\sum_{i=1}^K a_i(t)$. Note that a weight of the link $i$ can be obtained by $w_i \equiv \sum_{t=1}^T a_i(t)$. We also denote the number of newly activated links at the time step $t$ by $K(t)$. By definition $K(t)\leq W(t)$. The time series of $K(t)$ and $W(t)$ are written as $\{K(t)\}$ and $\{W(t)\}$, respectively, where $\sum_{t=1}^{T}K(t)=K$ and $\sum_{t=1}^{T}W(t)=W$. Figure~\ref{fig1} shows an example of $\{K(t)\}$ and $\{W(t)\}$ for the temporal network with $N=5$, $K=8$, $W=12$, and $T=9$. The time-resolved and time-aggregated representations of the temporal network are shown in the top panels, while $\{K(t)\}$ and $\{W(t)\}$ are presented in the bottom panel. 

Activity patterns in many temporal network datasets are known to be non-Poissonian or bursty~\cite{Karsai2018}, implying the existence of temporal correlations or memory effects, which are often found in multiple timescales. For characterizing the activity patterns with multiscale temporal correlations, we adopt the MSE method~\cite{Costa2002, Costa2005} among others such as Hurst exponent~\cite{Hurst1951}. It is because the face-to-face interaction datasets to be studied in our work have relatively short observation periods and the MSE method has been known to be less dependent on the time series length such as in physiological systems~\cite{Costa2002}.

\subsection{Multiscale entropy}
\label{MSE}
In this Subsection, we present a brief description of the multiscale entropy (MSE) method~\cite{Costa2002, Costa2005} for characterizing the time series with multiscale temporal correlations. We first define the sample entropy~\cite{Richman2000}. Let us consider a univariate discrete time series $\{x_t\}$ for $t=1,\dots,T$,
from which we get $T-m$ vectors of length $m$, i.e., $X_t^m=(x_t,\dots,x_{t+m-1})$ 
for $t=1,\dots,T-m$. For a given vector $X_t^m$, one can calculate the probability $C^m_t(r)$ that a random vector $X_{t'}^m$ for $t'\neq t$ lies within a distance $r$ from $X_t^m$, namely, satisfying $\max\{|x_{t+s}-x_{t'+s}|\}_{s=0,\dots,m-1} \leq r$. Then the average of $C_t^m(r)$ over $t$ is denoted by $U^m(r)\equiv (T-m)^{-1} \sum_{t=1}^{T-m}C_t^m(r)$. Similarly, one can get $T-m$ vectors of length $m+1$, denoted by $X_t^{m+1}$, from which $C_t^{m+1}(r)$ and $U^{m+1}(r)$ are respectively calculated. Using $U^m(r)$ and $U^{m+1}(r)$ one defines the sample entropy, denoted by $S_E(m,r)$, as follows:
\begin{align}
S_E(m,r)\equiv-\ln{\frac{U^{m+1}(r)}{U^m(r)}}=\ln{\frac{\sum_{t=1}^{T-m}C_t^m(r)}{\sum_{t=1}^{T-m}C_t^{m+1}(r)}}.
\label{eq:SE}
\end{align}
It is straightforward to see that the more random or complex time series tends to have the higher value of $S_E$. 
\begin{figure}[!t]
\center
\includegraphics[width=\columnwidth]{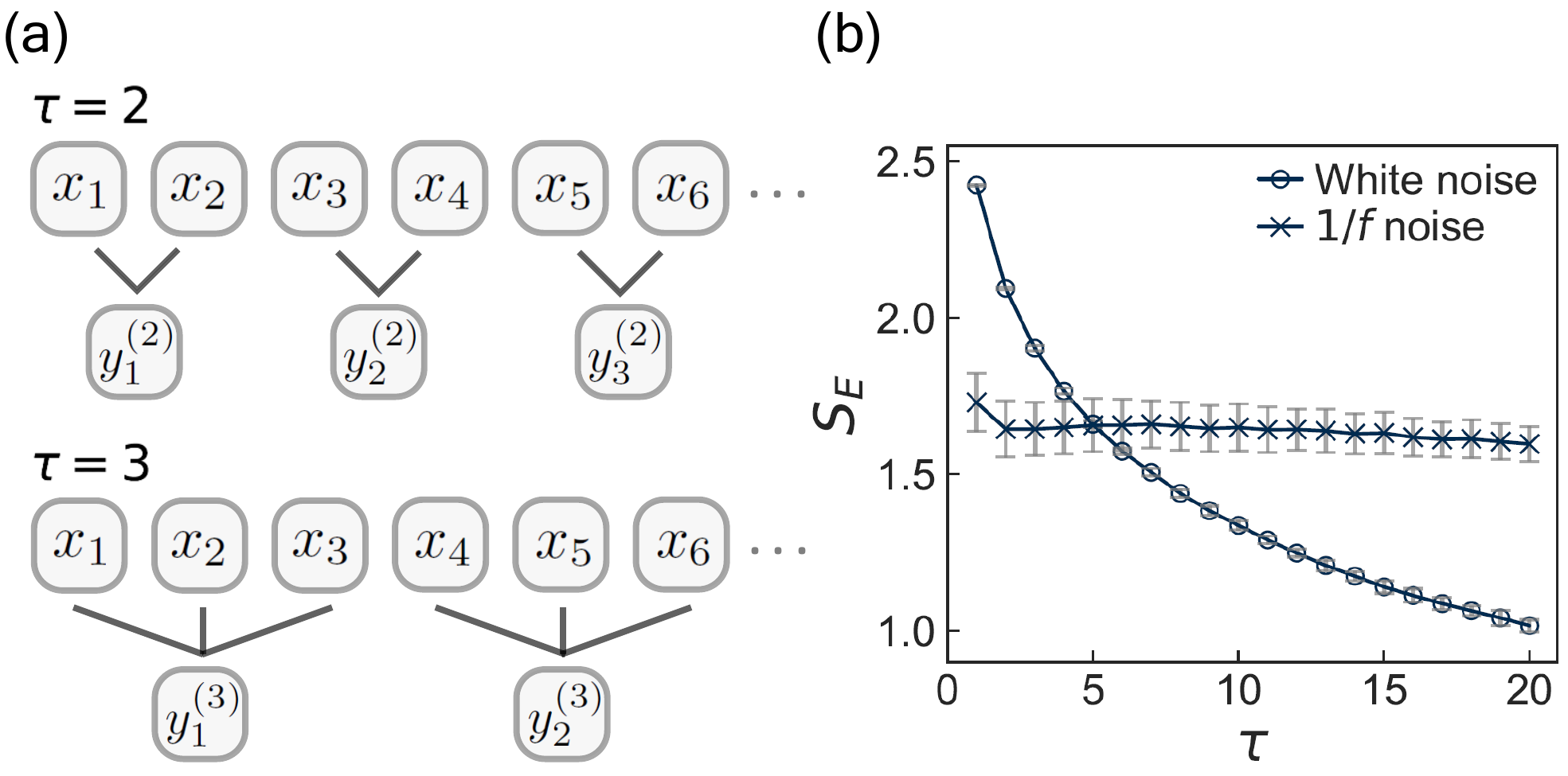}
\caption{Multiscale entropy (MSE) method: (a)~a schematic illustration of coarse-graining procedure of the time series $\{x_t\}$ for $t=1,\dots,T$ with the scale factor $\tau=2$ (top) and $3$ (bottom), and (b)~numerical results of the MSE method for the white noise and the $1/f$ noise. Each value of $S_E$ is averaged over $10$ time series with $T=3\times10^4$, and the error bar denotes its standard deviation.}
\label{fig2}
\end{figure}

To analyze the time series with multiscale temporal correlations by means of the sample entropy $S_E$, Costa~et~al. proposed the MSE method~\cite{Costa2002} by incorporating a coarse-graining procedure. For a given time series $\{x_t\}$ for $t=1,\dots,T$
and the scale factor $\tau$, the coarse-grained time series
$\{y_t^{(\tau)}\}$ for $t=1,\dots,T/\tau$
is constructed by averaging the elements in $\{x_t\}$ within non-overlapping time windows of size $\tau$ such that
\begin{align}
y_t^{(\tau)}=\frac{1}{\tau} \sum_{t'=(t-1)\tau+1}^{t\tau}x_{t'}\ \textrm{for}\ t=1,\dots,\frac{T}{\tau},
\end{align}
see Fig.~\ref{fig2}(a). If $\tau=1$, the time series $\{y_t^{(1)}\}$ is the same as $\{x_t\}$. The sample entropy $S_E$ of $\{y_t^{(\tau)}\}$ is calculated for various values of $\tau$, which is called the MSE method. This method has been applied to various datasets, e.g., for heartbeat, neural, and atmospheric time series~\cite{Costa2002, Costa2005, Liu2015, Nogueira2017}. For the rest of the paper, we will calculate $S_E$ in Eq.~\eqref{eq:SE} using $m=2$ and $r=0.15\sigma$ with $\sigma$ denoting the standard deviation of $\{x_t\}$ in all cases.

For the demonstration of the MSE method, we apply this method to two kinds of time series, i.e., white noise and $1/f$ noise, as done in Refs.~\cite{Costa2002,Costa2005}. The white noise is uncorrelated time series, while the $1/f$ noise is known to have long-range temporal correlations. The results of the MSE method for these time series are presented in Fig.~\ref{fig2}(b).  
In the case of the white noise, $S_E$ monotonically decreases as $\tau$ increases, whereas $S_E$ remains almost constant for the $1/f$ noise. The monotonically decreasing $S_E$ for the white noise implies that the noisy, random behavior tends to be averaged out for the larger scale factor. In contrast, the overall constant $S_E$ for the $1/f$ noise indicates the temporal self-similarity due to the long-range temporal correlations.

\subsection{Empirical results for temporal networks}
\label{empirical results}
We consider six empirical face-to-face interaction datasets provided by the SocioPatterns project~\cite{SocioWeb}: a primary school dataset for 2 days ($N=242$), a hospital dataset for 5 days from 6 a.m. to 8 p.m. for each day ($N=75$), a workplace dataset for 10 days ($N=92$), a high school dataset for 4 days in 2011 ($N=126$), a high school dataset for 7 days in 2012 ($N=180$), and a conference dataset for 3 days ($N=113$). The high school datasets in 2011 and 2012 are denoted as ``school (2011)'' and ``school (2012)'', respectively. In all datasets, contacts or interactions between individuals were recorded every 20 seconds, defining the unit of the time step in our work. 

\begin{figure}[!t]
\center
\includegraphics[width=\columnwidth]{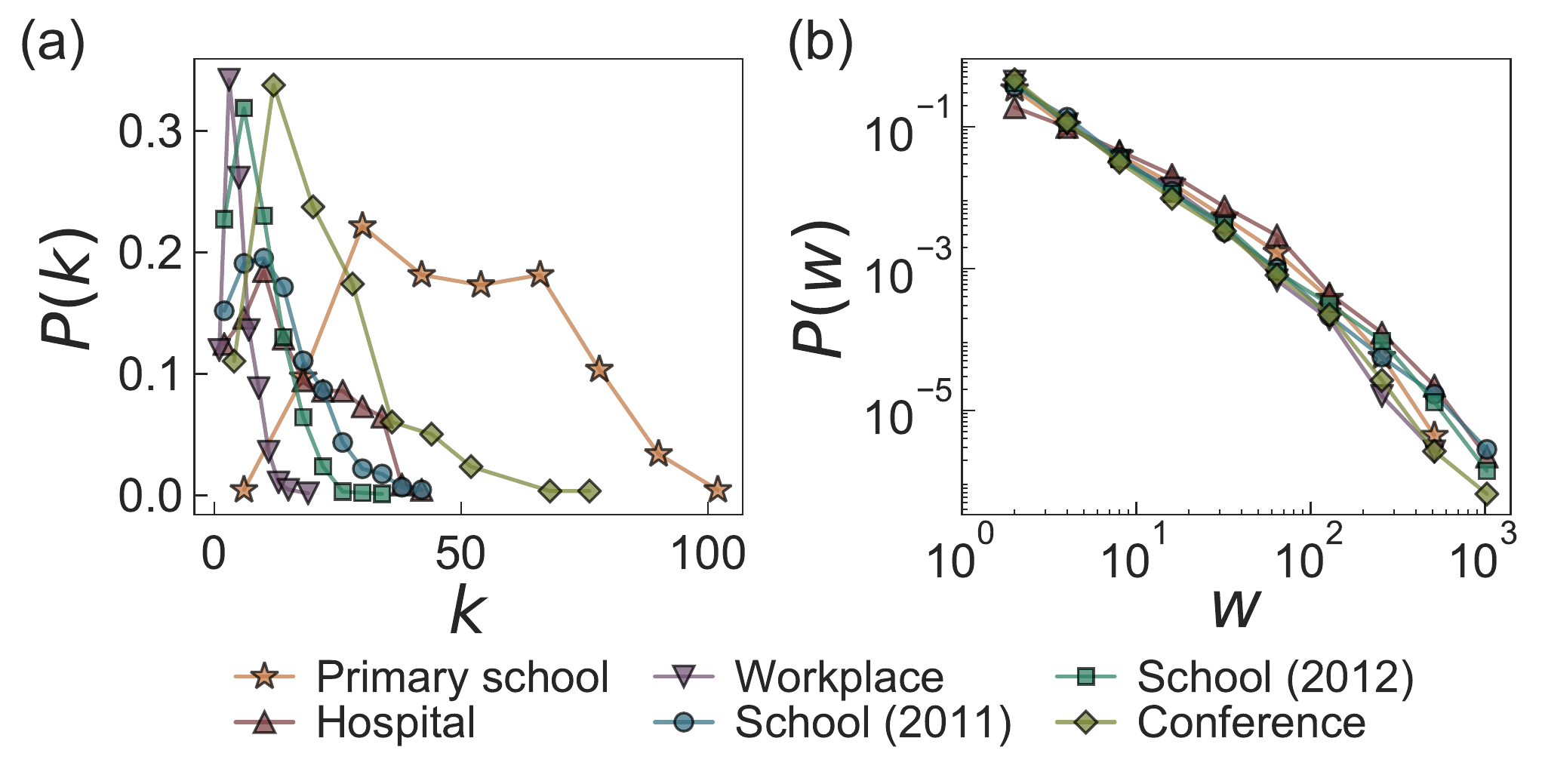}
\caption{Degree distributions $P(k)$ (a) and weight distributions $P(w)$ (b) of time-aggregated networks for six face-to-face interaction datasets, i.e., for the primary school (\ding{73}), hospital ($\triangle$), workplace ($\triangledown$), school (2011) ($\bigcirc$), school (2012) ($\square$), and conference ($\Diamond$). $P(k)$s are linearly binned, while $P(w)$s are logarithmically binned.}
\label{fig3}
\end{figure}

We first investigate the basic topological properties of time-aggregated networks. We obtain the time-aggregated network for each day of each dataset to get degree and weight distributions, where the degree $k$ means the number of neighbors. These daily distributions are averaged for each dataset to get the averaged $P(k)$ and $P(w)$, as shown in Fig.~\ref{fig3}. We observe that $P(k)$s show increasing and then decreasing behaviors, while being mostly right-skewed, except for the case with the primary school. $P(k)$ for the primary school shows both large average and large variance of degrees. The weight distributions $P(w)$ are found to show the similar heavy-tailed behaviors across all datasets. We conclude that the topological structures of time-aggregated networks of six datasets are qualitatively similar to each other.

\begin{figure*}[!t]
\center
\includegraphics[width=\textwidth]{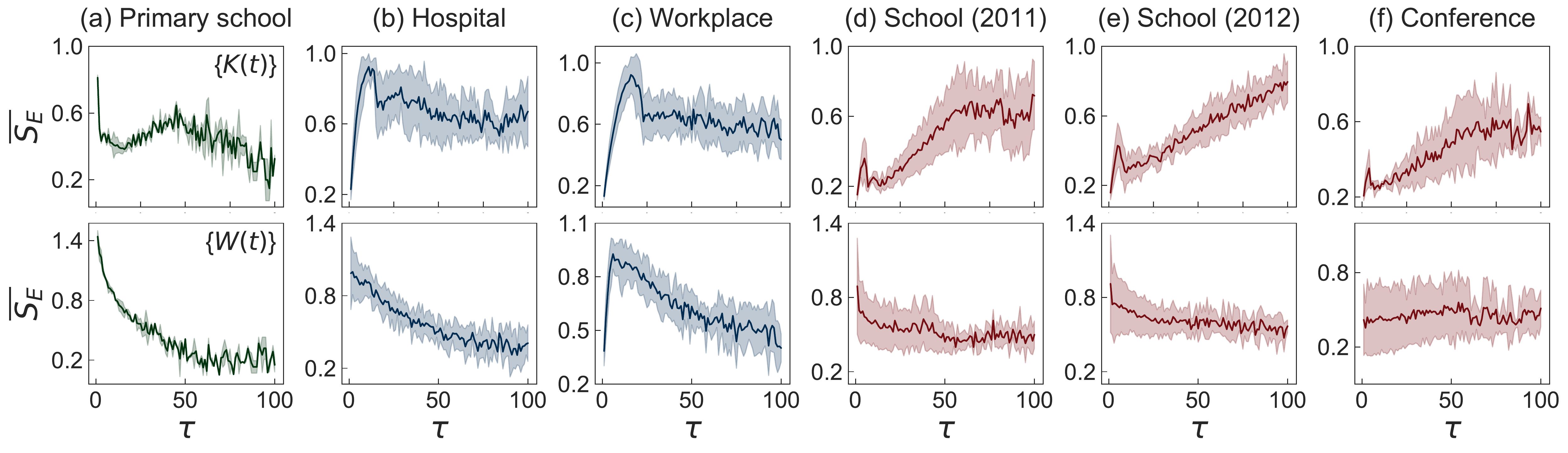}
\caption{Results of the MSE method applied to time series $\{ K(t)\}$ (top panels) and $\{W(t)\}$ (bottom panels) of six face-to-face interaction datasets for (a)~primary school, (b)~hospital, (c)~workplace, (d)~school (2011), (e)~school (2012), and (f)~conference. See the main text for details.}
\label{fig4}
\end{figure*}

Next, we apply the MSE method to the time series $\{K(t)\}$ and $\{W(t)\}$ derived from the above mentioned datasets. For each day of each dataset, we calculate the sample entropy $S_E$ for the coarse-grained time series using the scale factor of $\tau=1,\dots,100$. 
Then the curves of $S_E$ as a function of $\tau$ are averaged over all days for each dataset, denoted by $\overline{S_E}$. The results of $\overline{S_E}$ are presented with the corresponding standard deviations in Fig.~\ref{fig4}, where the top (bottom) panels show the results for $\{K(t)\}$ ($\{W(t)\}$).

According to the behavioral patterns of $\overline{S_E}$ for $\{K(t)\}$, the six datasets can be divided into three categories: (i) The primary school dataset shows the overall increasing and then decreasing behavior of $\overline{S_E}$, apart from the peak at $\tau=1$ [top panel in Fig.~\ref{fig4}(a)]. Note that the decreasing behavior of $\overline{S_E}$ was observed for the white noise in Fig.~\ref{fig2}(b). (ii) $\overline{S_E}$ for hospital and workplace datasets increases quickly and then decreases very slowly or even fluctuates around some constant [top panels in Fig.~\ref{fig4}(b,~c)]. This fluctuating behavior is similar to the result for the $1/f$ noise in Fig.~\ref{fig2}(b), implying the long-range temporal correlations. (iii) The other three datasets, i.e., school (2011), school (2012), and conference, show the overall increasing $\overline{S_E}$ [top panels in Fig.~\ref{fig4}(d--f)], indicating that the time series appears to be more random or complex when looked at in longer timescales. The similar increasing behaviors have been reported for neural time series~\cite{Misic2010, Misic2011, Bosl2011}.

Results for $\{W(t)\}$ in the bottom panels of Fig.~\ref{fig4} can be better understood by comparing them with those for $\{K(t)\}$ as $W(t)$ is the sum of $K(t)$ and the number of activated links that have been activated before the time $t$. The latter kind of activations, corresponding to $W(t)-K(t)$, indeed leads to different behaviors of $\overline{S_E}$ for $\{W(t)\}$ than those for $\{K(t)\}$: In the case with the primary school, $\overline{S_E}$ for $\{W(t)\}$ overall monotonically decreases, implying that the values of $\{W(t)\}$ are more uncorrelated with each other than those of $\{K(t)\}$. $\overline{S_E}$ for the hospital and workplace datasets overall decreases for the almost entire range of $\tau$, implying the long-range correlations in $\{K(t)\}$ must have been largely destroyed in $\{W(t)\}$. Finally, the other three datasets for school (2011), school (2012), and conference show the almost flat behaviors of $\overline{S_E}$ for $\{W(t)\}$, similarly to the case with $1/f$ noise. In sum, we find that the activations observed by $W(t)-K(t)$ tend to weaken the temporal correlations present in $\{K(t)\}$.

\begin{figure}[!t]
\center
\includegraphics[width=\columnwidth]{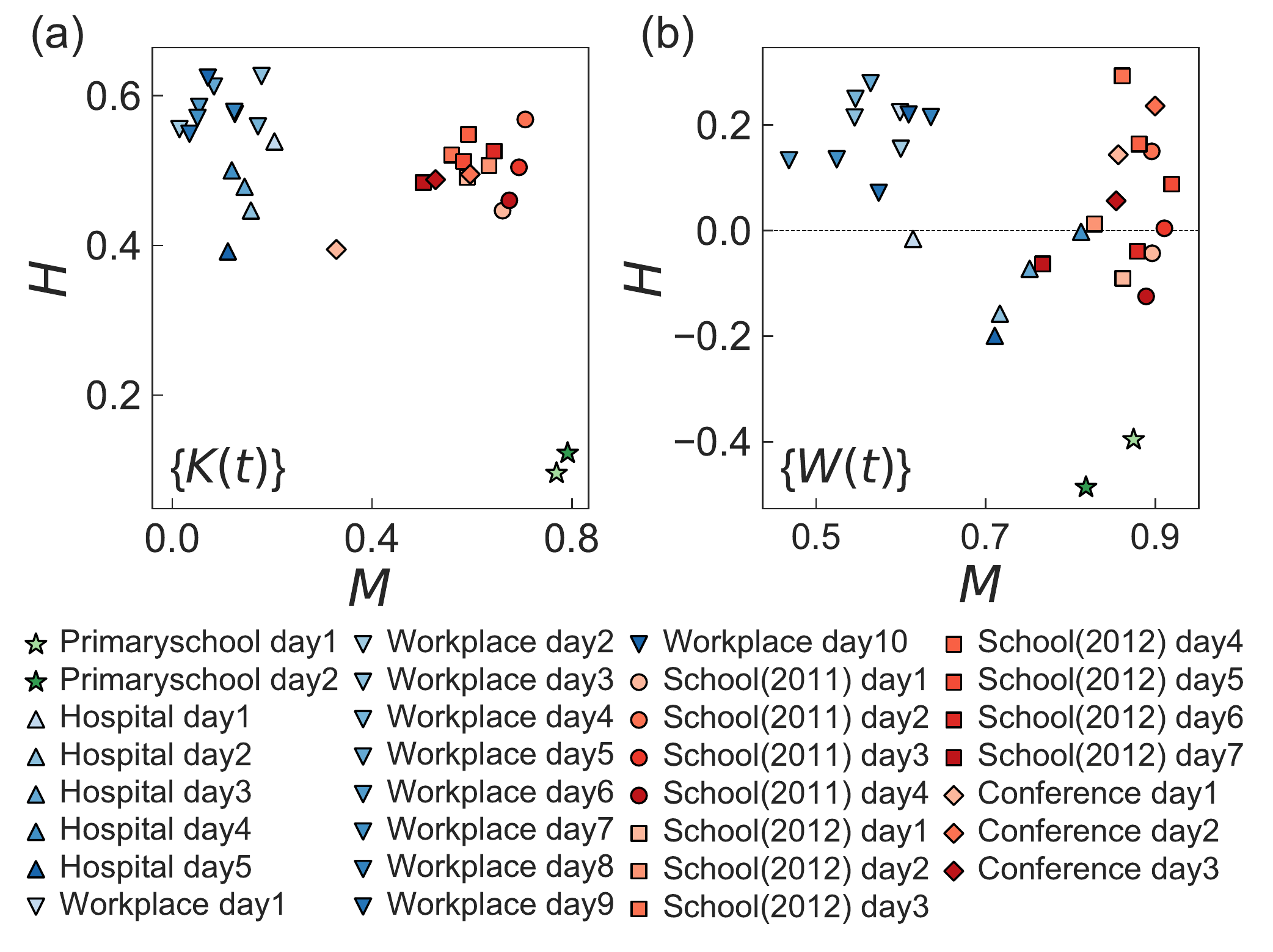}
\caption{Scatter plot of values of $H$ in Eq.~\eqref{eq:H} and $M$ in Eq.~\eqref{eq:M} in the $(M,H)$-space for time series $\{K(t)\}$ (a) and $\{W(t)\}$ (b) using the same datasets analyzed in Fig.~\ref{fig4}.}
\label{fig5}
\end{figure}

For more detailed understanding of the empirical results by the MSE method, we introduce two quantities for characterizing $\{K(t)\}$ and $\{W(t)\}$: the heterogeneity level $H$ and the memory coefficient $M$. These quantities are based on the burstiness parameter and memory coefficient that were originally proposed in Ref.~\cite{Goh2008} for measuring the temporal correlations in the point processes in terms of interevent times. In our work, instead of interevent times, we analyze the values of time series of $\{x_t\}$ for $t=1,\dots,T$. To measure how broad the distribution of values of $x_t$ is compared to their mean, we calculate the mean and standard deviation of the values of $x_t$, respectively denoted by $m_x$ and $\sigma_x$, to define the heterogeneity level $H$ as follows:
\begin{align}
H\equiv \frac{\sigma_{x}-m_{x}}{\sigma_{x}+m_{x}}.
\label{eq:H}
\end{align}
If all values of $x(t)$ are the same, one gets $H=-1$, while $H=0$ in the case when $x_t$ is exponentially distributed. If the distribution of $x_t$ is heavy tailed, $H>0$ is expected. The memory coefficient $M$ for the time series of $\{x_t\}$ is defined as
\begin{align}
M\equiv \frac{1}{T-1}\sum_{t=1}^{T-1}\frac{(x_t-m_1)(x_{t+1}-m_2)}{\sigma_1\sigma_2},
\label{eq:M}
\end{align}
where $m_{1}$ and $\sigma_{1}$ ($m_{2}$ and $\sigma_{2}$) are the mean and standard deviation of $\{x_1,\dots,x_{T-1}\}$ ($\{x_2,\dots,x_{T}\}$), respectively. The value of $M$ ranges from $-1$ to $1$. If a large (small) $x_t$ tends to be followed by the large (small) $x_{t+1}$, $M$ is positive, while $M$ is negative in the opposite case.

We calculate the values of $H$ and $M$ for $\{K(t)\}$ and $\{W(t)\}$ for each day of each dataset. These values are plotted in the $(M,H)$-spaces as shown in Fig.~\ref{fig5}. In the case with $\{K(t)\}$, we clearly find three clusters of points: (i) The primary school dataset is characterized by the smallest values of $H$ ($\approx0.1$) and the largest values of $M$ ($\approx0.8$), implying that the values of the time series are relatively homogeneous, while they are strongly correlated with each other. (ii) The hospital and workplace datasets show the large values of $H$ ($0.4\lesssim H\lesssim 0.7$) and the small values of $M$ ($0\lesssim M\lesssim 0.2$). It means that the values of the time series are highly heterogeneous, but showing with relatively weak correlations between them. (iii) The other three datasets, i.e., school (2011), school (2012), and conference, show large values of both $H$ and $M$ such that the values of the time series are highly heterogeneous as well as strongly correlated with each other. From the results for $\{W(t)\}$, we can observe three clusters similarly to those for $\{K(t)\}$, apart from the observation that the values of $H$ ($M$) are overall much smaller (larger) than those of $\{K(t)\}$.

We remark that three clusters identified in the $(M,H)$-spaces one-to-one correspond to three different behavioral patterns of $S_E$ as a function of $\tau$ as discussed above. From such a correspondence one can guess that heterogeneous values of the time series, i.e., large $H$, are necessary to show the non-decreasing behaviors of $S_E$. Further, the increasing $S_E$ could additionally require strong positive correlations between consecutive values of the time series.

Interestingly, the datasets in each cluster turn out to share similar social conditions either enhancing or suppressing interactions between individuals. In particular, we focus on the temporal behaviors of such conditions or environmental changes. The participants in the primary school dataset could have break times but only three times including lunch per day~\cite{Stehle2011}, while in the high school and conference cases, the interaction between participants were affected by scheduled programs with several breaks~\cite{Fournet2014, Isella2011}. During the breaks participants have chances to introduce each other or strengthen their existing relations, while such interactions can be relatively suppressed for the rest of the observation periods. Unlike schools and conference, there were no constrained schedules for the participants in the hospital and workplace datasets~\cite{Vanhems2013,Genois2015}. Generally speaking, the environmental changes can obviously influence the evolution of temporal networks, yet the effects of environmental changes on the evolution of temporal networks are far from being fully understood. To explore such effects, in the following Section we will devise and study a temporal network model that qualitatively reproduces the observed patterns by incorporating the environmental changes.

\begin{figure}[!t]
\center
\includegraphics[width=\columnwidth]{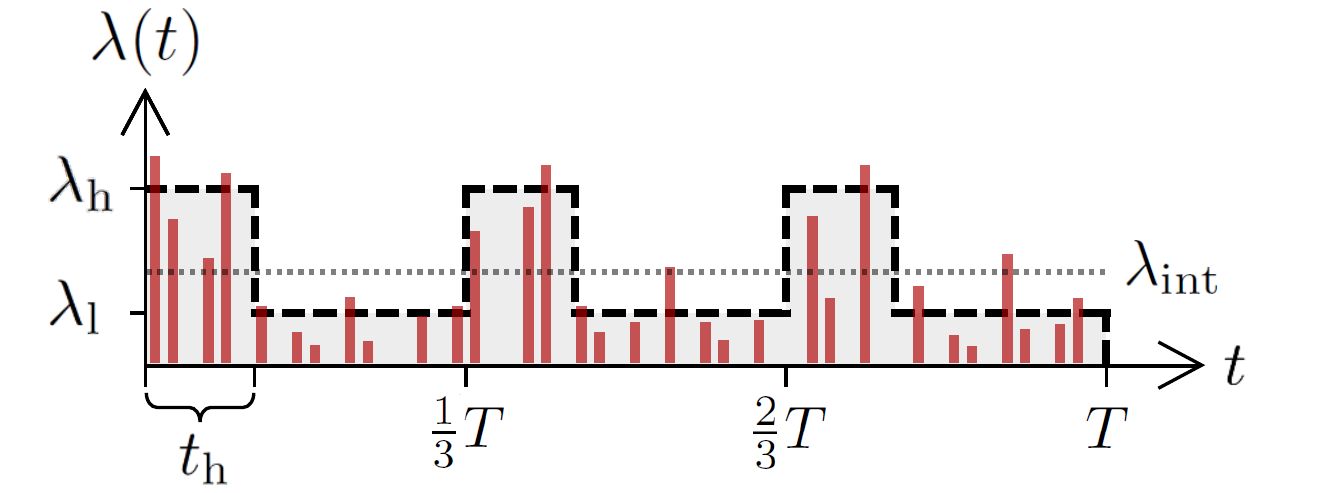}
\caption{An illustration of the model for generating a periodic time series $\{z(t)\}$ for $t=1,\dots,T$ (red vertical lines) using Eqs.~\eqref{zt} and~\eqref{P_zt}, where the time-varying parameter $\lambda(t)$ in Eq.~\eqref{P_zt} (thick dashed curve) is shaped by three parameters, i.e., $\lambda_{\rm h}$, $\lambda_{\rm l}$, and $t_{\rm h}$, in the case with $n=3$. The horizontal dotted line for $\lambda_{\rm int}$ is plotted for comparison. See the main text for details.}
\label{fig6}
\end{figure}

\section{Temporal network model}
\label{section3}

To explore the impact of environmental changes on the dynamics of temporal networks, we will first investigate a model for generating a periodic time series by considering both external and internal factors. This time series could represent either $\{K(t)\}$ or $\{L(t)\}$, where $L(t)\equiv W(t)-K(t)$ denotes the number of links that have previously been activated and are activated at the time step $t$ as well. Then, based on the periodic time series model, we will devise and study a temporal network model showing various temporal interaction patterns observed in the empirical datasets.

\subsection{Modeling a periodic time series}
\label{subsection3-1}

We devise a model for generating a periodic time series $\{z(t)\}$ for $t=1,\dots,T$, whose values are determined by both the external and internal factors. Considering the fact that the value of the time series of our interest is not always positive in the empirical analysis, we introduce the probability of having a positive $z(t)$, which is denoted by $\rho$ ($0<\rho<1$). Then one can write
\begin{align}
z(t)=
\begin{dcases}
~0 & \text{with}~1-\rho,\\
~l & \text{with}~\rho,
\end{dcases}
\label{zt}
\end{align}
where the positive integer $l$ is drawn from an exponential distribution $P(l;\lambda(t))$ with a time-varying parameter $\lambda(t)$, that is,
\begin{align}
P(l;\lambda(t))=\lambda(t)^{-1} e^{-l/\lambda(t)}.
\label{P_zt}
\end{align}
The time-varying parameter $\lambda(t)$ can be written as $\lambda_{\rm int}(t)+\lambda_{\rm ext}(t)$, where $\lambda_{\rm int}(t)$ and $\lambda_{\rm ext}(t)$ are the rates of spontaneous and externally-driven activations, respectively. We assume that $\lambda_{\rm int}(t)$ is constant of time, i.e., $\lambda_{\rm int}(t)=\lambda_{\rm int}$, while $\lambda_{\rm ext}(t)$ is a periodic function whose time average vanishes. The positive (negative) $\lambda_{\rm ext}(t)$ enhances (suppresses) activations compared to the constant activity level of $\lambda_{\rm int}$. 

For simplicity, we assume that $\lambda(t)$ has only two levels of activity, i.e., $\lambda_{\rm h}$ and $\lambda_{\rm l}$ ($\lambda_{\rm h}\geq \lambda_{\rm l}$). To be precise, the total period $T$ is divided into $n$ intervals. Each interval of length $T/n$ starts with a high activity period of length $t_{\rm h}$, for which $\lambda(t)=\lambda_{\rm h}$. This is followed by a low activity period of length $T/n-t_{\rm h}$, for which $\lambda(t)=\lambda_{\rm l}$. Note that $t_{\rm h}\leq T/n$. In Fig.~\ref{fig6} we present an example of $\lambda(t)$ for the case with $n=3$. The sum of $z(t)$ over the entire period of $T$ is assumed to be given as a control parameter $Z$, namely,
\begin{align}
Z=\sum_{t=1}^T z(t)=\rho\sum_{t=1}^{T}\lambda(t)=\rho\left[\lambda_{\rm h} t_{\rm h}+\lambda_{\rm l}\left(\frac{T}{n}-t_{\rm h}\right)\right]n,
\label{Z_condition}
\end{align}
leaving us with two independent parameters out of $\lambda_{\rm h}$, $\lambda_{\rm l}$, and $t_{\rm h}$, provided that $Z$, $T$, $\rho$, and $n$ are fixed. The external effect can be controlled mainly by the ratio $\lambda_{\rm h}/\lambda_{\rm l}$ and $t_{\rm h}$, where the larger ratio tends to be associated with the shorter period of $t_{\rm h}$. 
The case with $\lambda_{\rm h}/\lambda_{\rm l}=1$ implies no external effect ($\lambda_{\rm ext}(t)=0$), leading to $\lambda(t)=\lambda_{\rm int}=Z/(\rho T)$, which is also obtained when $t_{\rm h}=T/n$.

\begin{figure}[!t]
\includegraphics[width=\columnwidth]{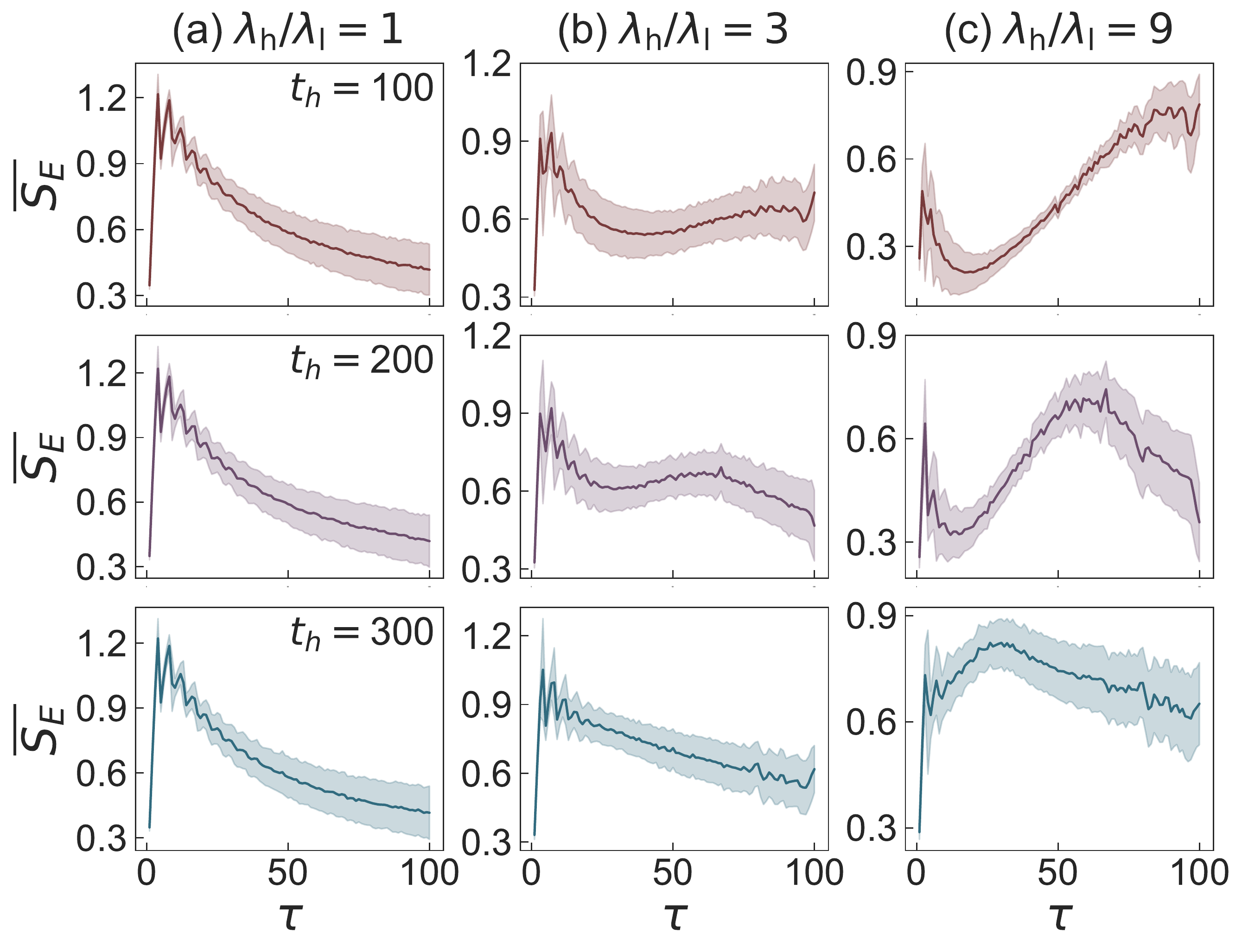}
\center 
\caption{Results of the MSE method applied to the time series $\{z(t)\}$ using the periodic time series model with $Z=1000$, $T=2000$, $\rho=0.2$, and $n=5$ for values of $\lambda_{\rm h}/\lambda_{\rm l}=1$, $3$, and $9$ (left to right) and $t_{\rm h}=100$, $200$, and $300$ (top to bottom). For each panel, we have generated $10^3$ time series to get the averaged curve $\overline{S_E}$ as a function of the scale factor $\tau$.}
\label{fig7}
\end{figure}

For each combination of $\lambda_{\rm h}/\lambda_{\rm l}$ and $t_{\rm h}$, we generate $10^3$ time series $\{z(t)\}$ with fixed values of $Z=1000$, $T=2000$, $\rho=0.2$, and $n=5$. The multiscale entropy (MSE) method is applied to each time series to get the averaged curve of $\overline{S_E}$ as a function of the scale factor $\tau$, as shown in Fig.~\ref{fig7}. In the case without external effect, i.e., $\lambda_{\rm h}/\lambda_{\rm l}=1$, we observe the overall decreasing behavior of $\overline{S_E}$, which was observed in the white noise [Fig.~\ref{fig2}(b)]. As expected, $t_{\rm h}$ has no effects on the results. As the periodic external effect gets stronger with the larger values of $\lambda_{\rm h}/\lambda_{\rm l}$, we find overall flat or even increasing behaviors of $\overline{S_E}$, as depicted in Fig.~\ref{fig7}(b,~c). Note that the overall flat behavior of $\overline{S_E}$ for $\lambda_{\rm h}/\lambda_{\rm l}=3$ and $t_{\rm h}=100$ was observed in the analysis of $1/f$ noise [Fig.~\ref{fig2}(b)]. Furthermore, it turns out that as $t_{\rm h}$ increases from $100$, the range of $\tau$ for the flat or increasing $\overline{S_E}$ shrinks and it is followed by the decreasing $\overline{S_E}$ for the large $\tau$ regime.

For understanding the effects of $t_{\rm h}$ on the MSE results in the general case with $\lambda_{\rm h}/\lambda_{\rm l}>1$, we calculate the fluctuation of $\lambda(t)$ as follows:
\begin{align}
\sigma^2_{\lambda}\equiv \frac{1}{T}\sum_{t=1}^T[\lambda(t)-\lambda_{\rm int}]^2= \frac{\lambda_{\rm int}^2(\lambda_{\rm h}/\lambda_{\rm l}-1)^2 s(1-s)}{[1+(\lambda_{\rm h}/\lambda_{\rm l}-1)s]^2},
\end{align}
where $s\equiv t_{\rm h}/(T/n)$. The fluctuation $\sigma^2_{\lambda}$ is found to be a decreasing function of $t_{\rm h}$ in the range of $t_{\rm h}\geq T/[n(\lambda_{\rm h}/\lambda_{\rm l}+1)]$. For our case with $\lambda_{\rm h}/\lambda_{\rm l}=3$ and $T/n=400$, the fluctuation $\sigma^2_{\lambda}$ decreases for $t_{\rm h}\geq 100$, implying that $\lambda(t)$ approaches the constant function, i.e., $\lambda(t)=\lambda_{\rm int}$. Hence the overall decreasing behavior of $\overline{S_E}$ is expected for large values of $t_{\rm h}$.

Our results for the periodic time series model enable us to get insight into the empirical findings, i.e., decreasing, flat, and/or increasing $\overline{S_E}$, from the temporal network datasets in the previous Section to a large extent. 

\subsection{Modeling temporal networks}
\label{subsection3-2}

Using the periodic time series model in Subsec.~\ref{subsection3-1}, we now devise a temporal network model that generates various temporal interaction patterns by considering both external and internal factors. We assume that the periodically changing environment affects not only the topological structure of the network, i.e., newly activated links, but also the activity patterns of links that have previously been activated. For the latter kind of activity patterns, we also incorporate a preferential activation mechanism for the heavy-tailed weight distributions in Fig.~\ref{fig3}(b), which is inspired by a preferential attachment mechanism accounting for the power-law degree distributions in scale-free networks~\cite{Barabasi1999}.

\begin{figure*}[!t]
\includegraphics[width=0.9\textwidth]{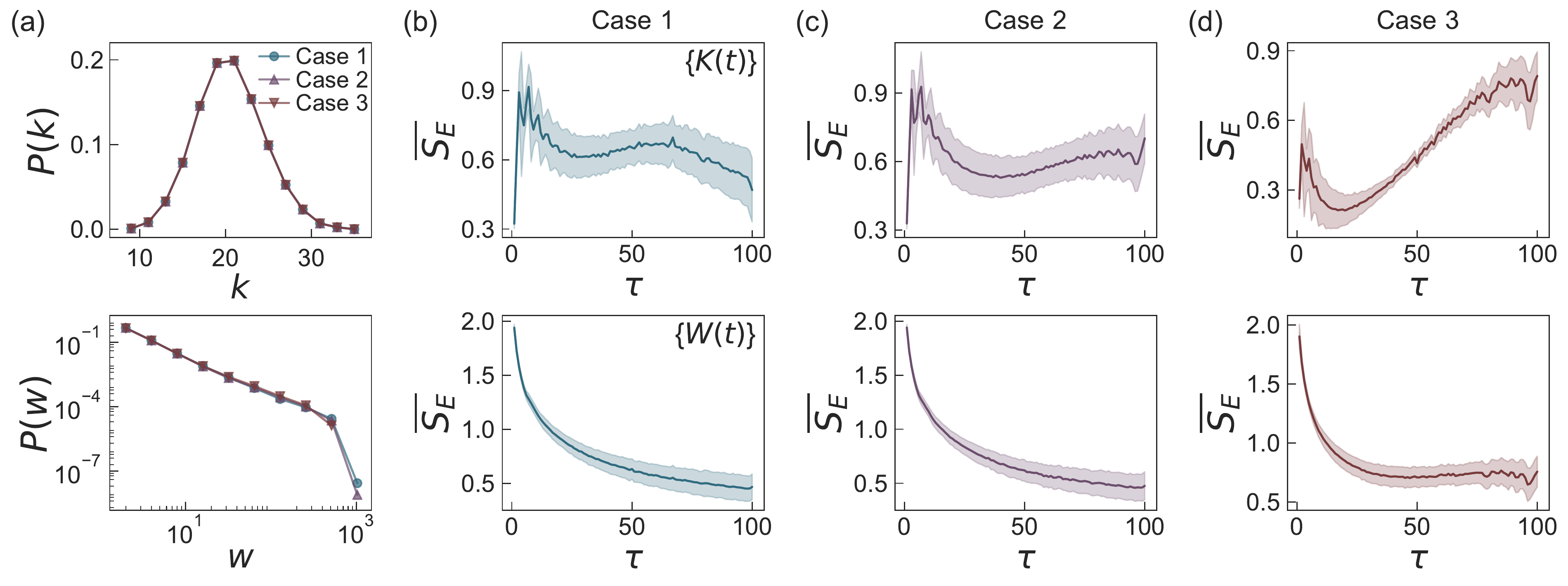}
\center
\caption{Simulation results of the temporal network model in terms of degree and weight distributions of time-aggregated networks (a) and the results by the MSE method applied to $\{K(t)\}$ (top panels) and $\{W(t)\}$ (bottom panels) (b--d). For each combination of parameter values $(t_{\rm h},\lambda_{K,\rm h}/\lambda_{K,\rm l},\lambda_{L,\rm h}/\lambda_{L,\rm l})=(200,3,1)$ (``Case 1''), $(100,3,1)$ (``Case 2''), and $(100,9,1.5)$ (``Case 3''), we generate $10^3$ temporal networks using the fixed values of $N=100$, $K=1000$, $W=10000$, $T=2000$, $\rho_K=0.2$, $\rho_L=0.8$, and $n=5$. In panel (a), degree distributions $P(k)$ are linearly binned, while weight distributions $P(w)$ are logarithmically binned.}
\label{fig8}
\end{figure*}

Based on the empirical degree distributions in Fig.~\ref{fig3}(a), we consider Erd\H{o}s-R\'{e}nyi (ER) random networks with $N$ nodes and $K$ links as substrate networks. That is, these $K$ links are topologically fixed but become activated and deactivated according to the activation rules. We begin with $K$ inactive links connecting $N$ nodes at the time step $t=0$. In the beginning of each time step $t$ for $t=1,\dots,T$, all links are divided into two groups: a set of links that have never been activated up to the time step $t-1$, denoted by $E_{K,t}$, and a set of links that have previously been activated up to the time step $t-1$, denoted by $E_{L,t}$. Then $K(t)$ links are uniformly chosen from $E_{K,t}$ to be activated for the first time, and $L(t)$ links are randomly chosen from $E_{L,t}$ according to the preferential activation mechanism (to be described) and then activated. Note that $|E_{L,t}|=\sum_{t'=1}^{t-1}K(t')$. Every activation lasts only for one time step before the next time step $t+1$ begins. The sum of $L(t)$ over the entire period of $T$ is denoted by $L$, defining the total number of activations across all links $W\equiv K+L$.

We assume that both $K(t)$ and $L(t)$ are affected by the periodically changing environment in a similar way. Therefore, the same periodic time series model in Subsec.~\ref{subsection3-1} can be used for both $K(t)$ and $L(t)$ but with different parameter values. Precisely, we use the symbols $\rho_K$, $\lambda_{K,\rm h}$, and $\lambda_{K,\rm l}$ ($\rho_L$, $\lambda_{L,\rm h}$, and $\lambda_{L,\rm l}$) for modeling $K(t)$ ($L(t)$), while $T$, $n$, and $t_{\rm h}$ have the same values for $K(t)$ and $L(t)$. These parameters should satisfy the following relations:
\begin{align}
K = \sum_{t=1}^T K(t) = \rho_K\left[\lambda_{K,\rm h} t_{\rm h}+\lambda_{K,\rm l}\left(\frac{T}{n}-t_{\rm h}\right)\right]n,
\label{K_condition}\\
L = \sum_{t=1}^T L(t) = \rho_L\left[\lambda_{L,\rm h} t_{\rm h}+\lambda_{L,\rm l}\left(\frac{T}{n}-t_{\rm h}\right)\right]n.
\label{L_condition}
\end{align}

As mentioned, $L(t)$ links are randomly chosen from $E_{L,t}$ according to the preferential activation mechanism, by which the more active links in the past are more likely to be activated in the future. The $L(t)$ links are chosen with probabilities proportional to their accumulated weights up to the time step $t-1$, i.e., 
\begin{align}
\label{prefer}
\Pi_i(t)=\frac{w_i(t-1)}{\sum_{j\in E_{L,t}} w_j(t-1)},
\end{align}
where $w_i(t)\equiv \sum_{t'=1}^{t}a_i(t')$. In the early stage of the simulation, $L(t)$ may exceed $|E_{L,t}|$, in which case a new random number is drawn from the distribution in a form of Eq.~\eqref{P_zt} until $L(t)\leq |E_{L,t}|$ is satisfied. This preferential activation mechanism is expected to result in the heavy-tailed weight distributions in the time-aggregated networks. Finally, $W(t)$ is given as $K(t)+L(t)$ at each time step $t$.

\subsection{Role of external effect in temporal networks}

We generate temporal networks using our temporal network model with the fixed values of 
$N=100$, $K=1000$, $W=10000$, $T=2000$, $\rho_K=0.2$, $\rho_L=0.8$, and $n=5$, but for various combinations of $t_{\rm h}$, $\lambda_{K,\rm h}/\lambda_{K,\rm l}$, and $\lambda_{L,\rm h}/\lambda_{L,\rm l}$. In particular, we consider three cases with parameter values of $(t_{\rm h},\lambda_{K,\rm h}/\lambda_{K,\rm l},\lambda_{L,\rm h}/\lambda_{L,\rm l})=(200,3,1)$ (``Case 1''), $(100,3,1)$ (``Case 2''), and $(100,9,1.5)$ (``Case 3''). These cases correspond to three categories identified by the empirical analysis of six face-to-face datasets in Sec.~\ref{section2}: Case 1 is for the primary school dataset, Case 2 is for the hospital and workplace datasets, and Case 3 is for the school (2011), school (2012), and conference datasets. For each case, $10^3$ temporal networks are generated for analysis. 

From the generated temporal networks, we first measure the degree and weight distributions of the time-aggregated networks, as shown in Fig.~\ref{fig8}(a). For all cases, $P(k)$s are binomial distributions as expected in the ER random networks, and $P(w)$s show heavy tails due to the preferential activation mechanism. Then, by applying the MSE method to the time series of $\{K(t)\}$ and $\{W(t)\}$, we calculate the averaged $\overline{S_E}$ with its standard deviation in Fig.~\ref{fig8}(b--d). It turns out that our temporal network model successfully generates various temporal interaction patterns observed in the empirical datasets using the above parameter values of $(t_{\rm h},\lambda_{K,\rm h}/\lambda_{K,\rm l},\lambda_{L,\rm h}/\lambda_{L,\rm l})$. For example, in Fig.~\ref{fig8}(d), $\overline{S_E}$ for $\{K(t)\}$ ($\{W(t)\}$) shows overall increasing (flat) behaviors, which have been observed in the analysis of datasets for school (2011), school (2012), and conference [see Fig.~\ref{fig4}(d--f)].

By the numerical simulation of our temporal network model, we have shown how the periodic external factor, when combined with the internal factor and the preferential activation mechanism, can induce complex temporal correlations in the network-level interaction patterns over a wide range of timescales. 

Finally, we remark that in our model the links are created (or activated for the first time) in different times, which may introduce some aging effects to the dynamics of temporal networks, e.g., as discussed in Ref.~\cite{Moinet2015}. The different creation times of links can also affect the dynamical processes taking place in temporal networks such as spreading~\cite{Holme2014, Holme2016}. In this sense our results highlight the need to study the impact of the environmental changes on the dynamics of temporal networks.

\section{Conclusion}
\label{section4}

The impact of environmental changes on the dynamics of temporal networks has been widely recognized, yet its understanding is far from complete. In our work we have analyzed six face-to-face interaction datasets in the framework of temporal networks by applying the multiscale entropy (MSE) method to the network-level time series. Based on the MSE results, we find that the temporal interaction patterns in those datasets can be categorized according to the environmental similarity, such as similar patterns of classes or break times in schools. To investigate the effects of periodic external factors on the various temporal interaction patterns, we first devise a model for generating a periodic time series to show that our model can reproduce various behaviors of the MSE results. Then we devise a temporal network model, based on the periodic time series model, that successfully generates various temporal interaction patterns in temporal networks. We also incorporate the preferential activation mechanism to account for the heavy-tailed distributions of link weights.

Our results demonstrate the importance of the environmental factors in understanding the dynamics of temporal networks. In particular, one can further investigate the possibilities of classifying the datasets according to the environmental similarity by applying our analysis method to other temporal network datasets. In addition, we have studied a temporal network model mainly focusing on the periodic external factors with only two levels of activity and on the random networks, while our model can be extended to take into account more realistic features such as complex network topologies (heavy-tailed degree distribution and/or community structure) and more realistic cyclic behaviors of environmental changes.

\begin{acknowledgments}
H.-H.J. acknowledges financial support by Basic Science Research Program through the National Research Foundation of Korea (NRF) grant funded by the Ministry of Education (NRF-2018R1D1A1A09081919).
H.J. acknowledges financial support by Basic Science Research Program through the National Research Foundation of Korea (NRF) grant funded by the Ministry of Education (NRF-2017R1A2B3006930).
\end{acknowledgments}

%
%

\end{document}